# On relation between separable indirect effect, natural indirect effect, and interventional indirect effect

Yan-Lin Chen[1], Sheng-Hsuan Lin[1,2,3,4*]

[1] Institute of Statistics, National Yang Ming Chiao Tung University, Hsinchu, Taiwan.
[2] Institute of Data Science and Engineering, National Yang Ming Chiao Tung University, Hsinchu, Taiwan.
[3] Department of Applied Mathematics, National Dong Hwa University, Hualien, Taiwan.
[4] Department of Biochemical and Molecular Medical Sciences, National Dong Hwa University, Hualien, Taiwan.

***Corresponding author**

Sheng-Hsuan Lin, MD, ScD
Institute of Statistics, National Yang Ming Chiao Tung University, Hsinchu, Taiwan.
1001 University Road, Hsinchu, Taiwan 300
Cell: +886 (3) 5712121 ext.56822
E-mail: shenglin@nycu.edu.tw

**Abstract:** Recently, the separable indirect effect (SIE) has gained attention due to its identifiability without requiring the untestable cross-world assumption necessary for the natural indirect effect (NIE). This article systematically compares the causal assumptions underlying the SIE, NIE, and interventional indirect effect (IIE) and evaluates their feasibility for mediational interpretation using the mediation null criterion, with a particular focus on the SIE. We demonstrate that, in the absence of intermediate confounders, the SIE lacks a mediational interpretation unless additional unverifiable assumptions are imposed. When intermediate confounders are present, separable effect methods fail to accurately capture the indirect effect, whereas the NIE still satisfy the mediation null criterion. Additionally, we present a new identification result for the NIE in the presence of intermediate confounders. Finally, we propose an integrated framework for practical analysis. This article emphasizes that the NIE is the most fundamental definition of indirect effect among the three measures and highlights the trade-off between mediational interpretability and assumption falsifiability.

# 1. INTRODUCTION

Causal mediation analysis is a specialized field focused on understanding how an exposure influences an outcome through an intermediate variable, known as a mediator [1]. A landmark concept in causal mediation analysis is the natural indirect effect (NIE)[2, 3], which has been widely recognized as the standard definition of indirect effect. However, when the identification assumptions of the NIE fail—such as in the presence of exposure-induced mediator-outcome confounders (referred to as intermediate confounders throughout this article)—the interventional indirect effect (IIE)[4], a randomized interventional analogue of the NIE, is often adopted as an alternative measure of indirect effect. In recent years, the concept of separable effects[5] has gained attention, particularly for defining causal effects in scenarios where the outcome of interest is subject to competing risks or truncation by death [6, 7]. The separable indirect effect (SIE) has also emerged as a popular alternative for mediation analysis. A comprehensive theoretical framework for the SIE, along with several applications in survival data, has been developed [8-10].

The motivation for the SIE in mediation analysis stems from Robins's critique of the limitations of the NIE: (a) the requirement that the mediator must be manipulable, (b) the difficulty in observing cross-world counterfactuals, and (c) the inability to experimentally falsify the assumptions required for its identification. To address these concerns, the SIE assumes that the exposure can be decomposed into two components: one affecting the mediator only and another affecting the outcome directly, without passing through the mediator. The SIE represents the effect of the first component while keeping the second component fixed. Unlike the NIE, the SIE does not require direct intervention on the mediator and is claimed to be verifiable in future experiments, provided researchers can successfully separate and randomly intervene on these exposure components [8]. The limitations of the NIE, particularly (a) and (b), align with Holland's famous principle "No causation without manipulation [11]." On the other hand, Pearl [12] argued that causal relationships exist independently of our ability to manipulate them, a perspective also supported by VanderWeele and Robinson [13]. This led

Pearl [12] to introduce the counter-slogan "Causation without manipulation? You bet!" Ideally, the IIE [4, 14, 15] can be validated through randomized controlled trials [16, 17], thereby circumventing (b) and (c), the challenges associated with the NIE. However, the IIE still cannot overcome (a), as it requires the mediator to be manipulable. From this perspective, the SIE may appear superior to both, but its validity as an indirect effect measure remains questionable and has not been thoroughly investigated.

Given the fundamental differences among the SIE, NIE, and IIE, a systematic comparison of these approaches is essential. This study explores the theoretical properties and underlying principles of the SIE by addressing the following key questions: (a) Under what assumptions and causal models does the SIE have a valid mediational interpretation? (b) How do these assumptions compare in strength to those of the NIE and IIE? To answer these questions, Section 2 introduces the definitions of the SIE, NIE, and IIE, along with the identification assumptions in cases without intermediate confounders. We then examine how these assumptions influence the validity of the mediational interpretation. Section 3 investigates the correspondence between path decomposition strategies of the separable effects and the path-specific effects defined by VanderWeele, Vansteelandt, and Robins [4]. Additionally, we propose a new identification result for the NIE, which remains valid even in the presence of intermediate confounders. Section 4 presents an integrated framework for practical analysis, and Section 5 summarizes our conclusions and future works.

## 2. CAUSAL MEDIATION ANALYSIS WITHOUT INTERMEDIATE CONFOUNDERS

### 2.1 Notations and review of graphical causal models

In this subsection, we review two commonly used graphical causal models: the nonparametric structural equation model with independent errors (NPSEM-IE)[12] and the finest fully randomized causally interpretable structured tree graph (FFRCISTG) [18]. Let $A, M, Y$ denote the exposure, mediator, and outcome of interested, with respective supports $\Omega_A$, $\Omega_M$, $\Omega_Y$. Let $C \in \Omega_C$ represent the set of baseline confounders that may affect two or more variables among $A$, $M$, and $Y$. To define measures of causal effects, we use the counterfactual notation [2, 19] $M(a = a')$, $Y(a = a')$, and $Y(a = a', m = m')$, which are abbreviated as $M(a')$, $Y(a')$, and $Y(a', m')$, where $a' \in \Omega_A$ and $m' \in \Omega_M$. A counterfactual variable $W(v = v')$ represents the hypothetical value that $W$ would take if $V$ were set to $v'$. For instance, $Y(a', m')$ denotes the hypothetical value that $Y$ would take if $A$ were intervened to $a'$ and $M$ to $m'$. Throughout this article, we focus on a binary exposure $A \in \Omega_A = \{0,1\}$.

The causal structure among $A$, $M$, and $Y$ is illustrated by the directed acyclic graph (DAG) in Figure 1(a) However, to identify causal effects of interest, additional assumptions are required. Therefore, two causal models—NPSEM-IE and FFRCISTG, both associated with the DAG in Figure 1(a)—are introduced. The NPSEM-IE associated with Figure 1(a) is shown in Figure 1(b). This model represents the data-generating process using unspecified structural functions $g_A, g_M, g_Y$: $A = g_A(C, \epsilon_A)$, $M(a') = g_M(A = a', C, \epsilon_M)$, $Y(a', m') = g_Y(A = a', M = m', C, \epsilon_Y)$, where $\epsilon_A$, $\epsilon_M$ and $\epsilon_Y$ are mutually independent error terms. The FFRCISTG associated with Figure 1(a) is represented by a single-world intervention graph (SWIG) [20] in Figure 1(c). Unlike the NPSEM-IE, this model does not specify common structural equations for all individuals but instead assumes mutual independence: $A \perp\!\!\!\perp M(a') \perp\!\!\!\perp Y(a', m') \mid C$, corresponding to d-separation [21] in Figure 1(c). This implies the following assumptions:

**Assumption 1.** $M(a') \perp\!\!\!\perp A \mid C,\ \forall a' \in \Omega_A,$

**Assumption 2.** $Y(a', m') \perp\!\!\!\perp A \mid C,\ \forall a' \in \Omega_A, m' \in \Omega_M,$

**Assumption 3.** $Y(a', m') \perp\!\!\!\perp M(a') \mid A, C,\ \forall a' \in \Omega_A, m' \in \Omega_M.$

In contrast, the NPSEM-IE model in Figure 1(b) implies not only Assumptions 1–3 but also an additional cross-world independence assumption:

**Assumption 4.** $Y(a', m') \perp\!\!\!\perp M(a'') \mid C,\ \forall a', a'' \in \Omega_A, m' \in \Omega_M.$

Both models also incorporate composition assumptions [22] and consistency assumptions [23]. We assume the positivity assumptions [24], which are essential for identification. It is important to note that NPSEM-IE is a submodel of FFRCISTG—that is, all assumptions in FFRCISTG are also implied by NPSEM-IE, but not vice versa. Although Assumption 4 is often considered difficult to hold [25], both models are essential for identifying the NIE, IIE, and SIE, which are reviewed in the following subsections.

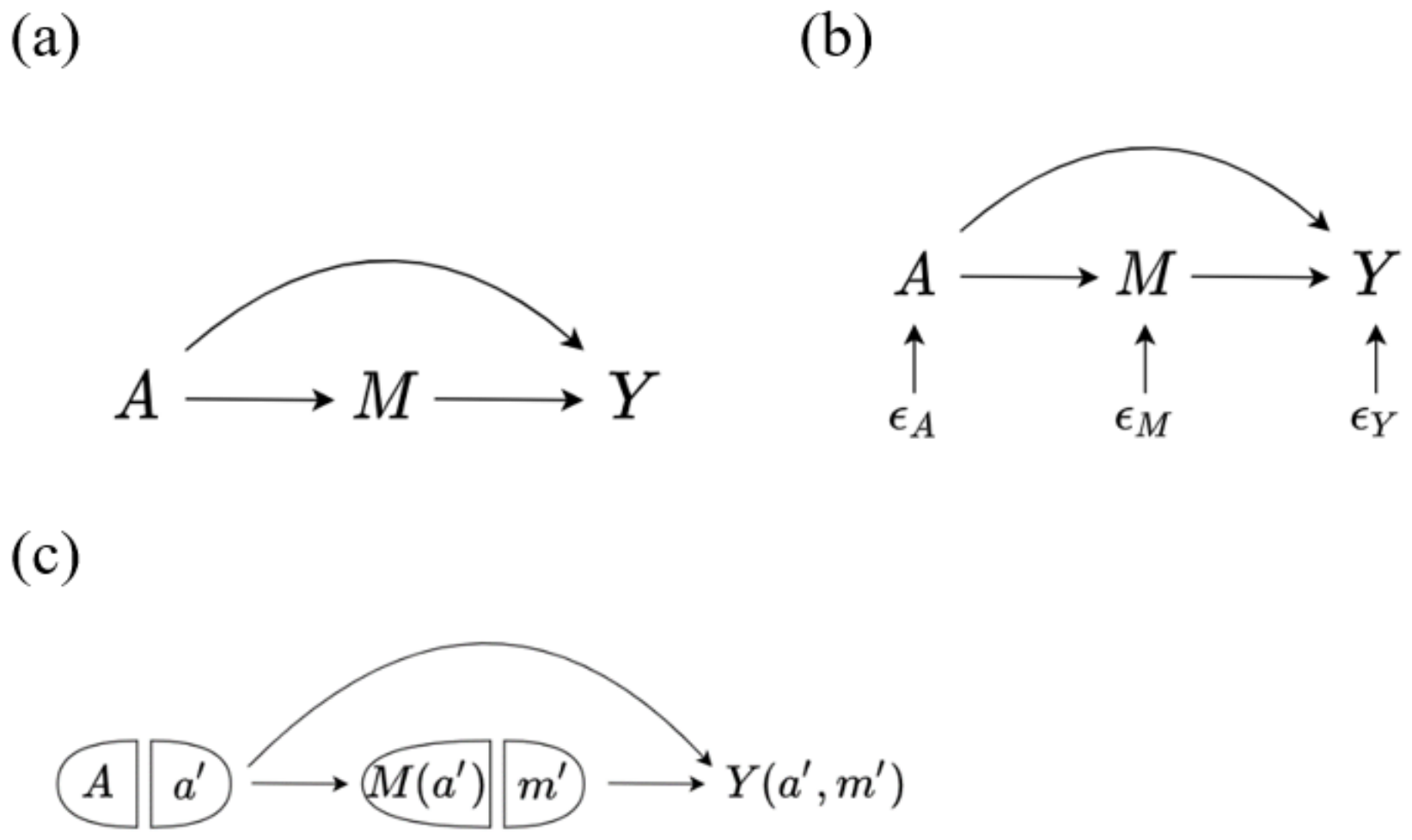

**Figure 1.** (a) A causal DAG for mediation analysis without intermediate confounders; (b) NPSEM-IE associated with Figure 1(a), where $\epsilon_A$, $\epsilon_M$ and $\epsilon_Y$ are assumed to be mutually independent error terms, which implies Assumptions 1-4 hold. (c) A template of SWIG constructed from Figure 1(a) which represents the independence assumption $A \perp\!\!\!\perp M(a') \perp\!\!\!\perp Y(a',m') \mid C$ of FFRCISTG. For simplicity, measurable baseline confounders $C$ among $A$, $M$ and $Y$ are omitted from the figures.

## 2.2 Review of natural effects, interventional effects, and their mediational interpretation

The NIE, first introduced by Robins and Greenland [2], is one of the most fundamental indirect effect measures used to explain how an exposure $A$ influences an outcome $Y$ through a mediator $M$. The natural effects method decomposes the total effect (TE), $\text{TE} \equiv E\{Y(1)\} - E\{Y(0)\}$, into the sum of the natural indirect effect (NIE) and the natural direct effect (NDE), which are defined as: $\text{NIE} \equiv E\{Y(1, M(1))\} - E\{Y(1, M(0))\}$, $\text{NDE} \equiv E\{Y(1, M(0))\} - E\{Y(0, M(0))\}$. The NIE captures the difference in outcomes when, under an intervention on the exposure, the mediator takes its natural value in the presence of exposure versus its natural value in the absence of exposure. The counterfactual variable $Y(a', M(a''))$ defined under this intervention strategy is referred to as a nested counterfactual. Under the NPSEM-IE in Figure 1(b), the NIE and NDE can be identified as:

**Formula 1.** $\text{NIE} \overset{id.}{=} Q(1,1) - Q(1,0)$,

**Formula 2.** $\text{NDE} \overset{id.}{=} Q(1,0) - Q(0,0)$,

where $Q(a', a'') = E[E\{E(Y|A = a', M, C)|A = a'', C\}]$ is referred to as the mediation formula.[12] In contrast, under the FFRCISTG in Figure 1(c), only upper and lower bounds [5] can be identified. However, as previously mentioned, the natural effects framework has key limitations. The nested counterfactual $Y(1, M(0))$ is cross-world, meaning it requires comparing counterfactual outcomes from different worlds, which cannot be observed in real-world experiments. Additionally, Assumption 4 in NPSEM-IE cannot be experimentally falsified, despite extensive discussions on alternative hypotheses [5, 25-27]. However, these alternative hypotheses also lack experimental falsifiability.

To address the limitations of the NIE, an interventional approach has been proposed as an alternative. In this framework [14, 15], the interventional indirect effect (IIE) and interventional direct effect (IDE) serve as substitutes for the NIE and NDE, respectively. They are defined as: $\text{IIE} \equiv E\{Y(1, G_C(1))\} - E\{Y(1, G_C(0))\}$, $\text{IDE} \equiv E\{Y(1, G_C(0))\} - E\{Y(0, G_C(0))\}$. Here $G_C(a')$ represents a random draw from $M(a')$ within strata defined by the same baseline $C$. Under the FFRCISTG model in Figure 1(c), both the IIE and IDE can be identified using the same formulas as the NIE and NDE (Formulas 1 and 2). Since FFRCISTG is a weaker model than NPSEM-IE, the IIE and IDE are also identifiable under NPSEM-IE. Miles [26] demonstrated that the IIE is identical to the NIE under the FFRCISTG model in Figure 1(c), provided an additional assumption—referred to as the no conditional mean causal interaction assumption [26]—holds:

**Assumption 5. (No Conditional Mean Causal Interaction)**

$$E\{Y(1, m') - Y(1, m'') - Y(0, m') + Y(0, m'')|M(0), C\} = 0, \forall m', m'' \in \Omega_M.$$

This assumption states that no causal interaction exists between exposure and the mediator on the outcome, at the population level, after conditioning on baseline confounders and the counterfactual value of the mediator in the absence of exposure. However, similar to Assumption 4 in NPSEM-IE, Assumption 5 is not experimentally falsifiable.

### 2.3 Review of separable effects

To achieve falsifiability, Robins and Richardson [5] introduced the SIE as a measure for indirect effect. This approach imposes additional constraints on the causal structure, as shown in Figure 2(a), which is a submodel of Figure 1(a). In this model, a key assumption is that exposure $A$ can be separated into two components—$A_M$ and $A_Y$. Here, $A_M$ represents the components of $A$ that affects $M$ but does not directly affect $Y$, while $A_Y$ represents the components of $A$ that affects $Y$ but not $M$. Notably, in the current dataset $A_M$ and $A_Y$ are determined by $A$. For simplicity, we abbreviate the counterfactuals $M(a_Y = a', a_M = a'')$ as $M_S(a', a'')$ and $Y(a_Y = a', a_M = a'')$ as $Y_S(a', a'')$. The separable indirect effect (SIE) and separable direct effect (SDE) are then defined as $\text{SIE} \equiv E\{Y_S(1,1)\} - E\{Y_S(1,0)\}$ and $\text{SDE} \equiv E\{Y_S(1,0)\} - E\{Y_S(0,0)\}$. Figure 2(a) assumes that there are no individual-level effects from $A_M$ to $Y$ or from $A_Y$ to $M$. This means that, for all individuals, the following individual-level isolation assumptions [28] hold:

**Assumption 6.** $Y(a_Y, a_M = a', m) = Y(a_Y, a_M = a'', m),\ \forall a', a'' \in \Omega_A,$

**Assumption 7.** $M(a_Y = a', a_M) = M(a_Y = a'', a_M),\ \forall a', a'' \in \Omega_A.$

Under these assumptions, the effect of the separated component $A_M$ on $Y$ can be intuitively interpreted as an indirect effect. Under the FFRCISTG model associated with Figure 2(a) (shown in Figure 2(c)), Assumptions 6 and 7 hold, allowing the SIE and SDE to be identified using Formula 1 and 2, respectively. In fact, even if Figure 2(a) is interpreted as an FFRCISTG at the population-level [8] (referred to as the "population FFRCISTG" illustrated in Figure 2(b)), the SIE remains identifiable using the same formulas, even without individual-level isolation assumptions. This model implies the following dismissible component assumptions

[6, 8]

**Assumption 8.**

$$E\{Y_S(a', a'')|M_S(a', a'') = m, C\} = E\{Y_S(a', a')|M_S(a', a') = m, C\},$$

**Assumption 9.** $P\{M_S(a', a'') = m'|C\} = P\{M_S(a'', a'') = m'|C\}$, $\forall m' \in \Omega_M$,

Additionally, the exchangeability assumptions (Assumptions 1–3) hold. Since $Y(a_Y = a', a_M = a'', m) \perp\!\!\!\perp M_S(a', a'') \,\big|\, C$, holds in the population FFRCISTG in Figure 2(b), for all $a', a'' \in \Omega_A$, Assumption 8 can be rewritten as $E\{Y(a_Y = a', a_M = a'', m')|C\} = E\{Y(a_Y = a', a_M = a', m')|C\}$, which ensures that there are no causal effects from $A_M$ to $Y$ at the population level. Similarly, Assumption 9 ensures that $A_Y$ does not affect $M$ at the population level.

It is important to note that the assumptions of the population FFRCISTG model in Figure 2(b) can be experimentally verified in future studies where components $A_M$ and $A_Y$ can be separately manipulated and randomly intervened. Moreover, Assumptions 1–3 are also, in principle, verifiable if the natural value of the mediator can be observed before intervention [8]. For instance, to assess Assumption 3 both $Y(a', m')$ and $M$ must be observed, meaning that one should observe $M$ before intervening on the mediator as $m'$.

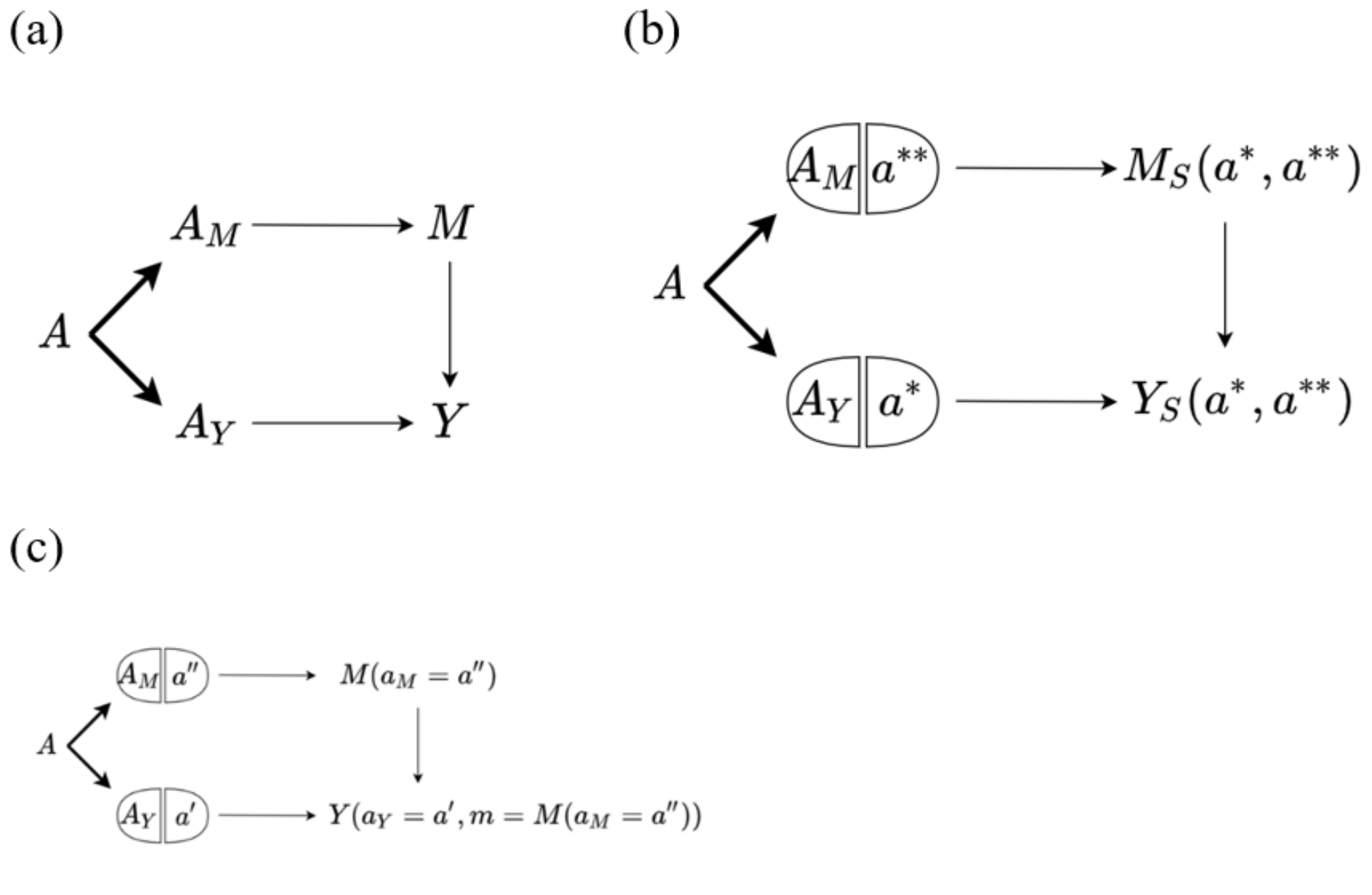

**Figure 2.** (a) A causal DAG extended from Figure 1(a), which is necessary for SIE; thick edges

indicate that $A_M$ and $A_Y$ are completely determined by $A$. (b) A population SWIG, representing the population FFRCISTG associated with Figure 2(a), which implies the dismissible component Assumptions 8 and 9 and the exchangeability Assumptions 1-3. Edges in the figure represent average causal effects at the population level. The absence of edges $a^* \rightarrow M_S(a^*, a^{**})$ and $a^{**} \rightarrow Y_S(a^*, a^{**})$ indicates the absence of population-level causal effects of $A_Y$ on $M$ and $A_M$ on $Y$. However, the form of individual counterfactual variables $M_S(a^*, a^{**})$ and $Y_S(a^*, a^{**})$ reveals that individual causal effects of $A_M$ on $Y$ and $A_Y$ on $M$ are still possible but should cancel out at the population level. (c) A SWIG representing the individual FFRCISTG associated with Figure 2(a), which implies the individual-level isolation Assumptions 6 and 7. For simplicity, measurable baseline confounders $C$ among $A$, $M$ and $Y$ are omitted from the figures.

### 2.4 Do the separable indirect effects have a mediational interpretation?

Although the identification assumptions for the SIE are verifiable, without a formal assessment, the SIE can only be interpreted as the effect of $A_M$ on $Y$. In this subsection, we examine whether the SIE has a valid mediational interpretation under the FFRCISTG model in Figure 2(c) and the population FFRCISTG model in Figure 2(b). To address this, we apply the sharper null criterion proposed by Miles [26], referred to as the mediation null criterion throughout this article:

**Definition (mediation null criterion).**

*An indirect effect measure should be null if, for each individual* $i$, *either* $M_i(a') = M_i(a'')$ *or* $Y_i(a', m') = Y_i(a', m'')$ *for all* $a', a'' \in \Omega_A$, *and* $m', m'' \in \Omega_M$.

Intuitively, if the causal pathway $A \rightarrow M \rightarrow Y$ does not exist for all individuals—meaning that either the edge from exposure to mediator is absent or the edge from mediator to outcome is absent (a condition referred to as the mediation null condition)—then a reasonable indirect effect measure should be null.

Robins, Richardson [8] noted that, under the causal structure shown in Figure 2(a), if the individual-level isolation assumptions hold, then the SIE and SDE are mathematically equivalent to the NIE and NDE, respectively. This equivalence arises because the causal parameters are equal in this setting: $Y_S(a', a'') = Y(a_Y = a', a_M = a'', m = M(a_Y = a', a_M = a'')) = Y(a_Y = a', a_M = a', m = M(a_Y = a'', a_M = a'')) = Y(a', M(a''))$. Consequently, we have the following theorem.

> **Theorem 1.** *The SIE,* $E\{Y_S(1,1) - Y_S(1,0)\}$*, satisfies the mediation null criterion under Assumptions 6-7.*

To explain this result, note that the mediation null condition ensures that for all individuals, $Y(1, M(1)) - Y(1, M(0)) = 0$, implying that the NIE equals zero and thus satisfies the mediation null criterion. Furthermore, since Assumptions 6 and 7 guarantee that the SIE is equal to the NIE, it follows that the SIE must also satisfy this criterion. However, the individual-level isolation assumptions are quite strong. Given this limitation, we instead examine whether the SIE retains its mediational interpretation under more empirically verifiable assumptions. However, under the weaker assumptions, the SIE lacks a mediational interpretation, as stated in the following theorem.

> **Theorem 2.** *The SIE,* $E\{Y_S(1,1) - Y_S(1,0)\}$*, does not satisfy the mediation null criterion under the population FFRCISTG in Figure 2(b).*

The proof of the theorem is provided in Supplementary Material A. The reason the SIE fails to satisfy the mediation null criterion is that, under the population FFRCISTG in Figure 2(b), it is identified as $E[\int_{m'} E(Y(1,m')|C)\{dF_{M(1)|C}(m'|C) - dF_{M(0)|C}(m'|C)\}]$. However, when there exists a cross-world confounder between $Y(1,m)$ and $M(0)$, which does not need to be controlled for under the FFRCISTG, this formula does not necessarily equal zero under the mediation null condition.

Miles [26] has demonstrated that the IIE does not satisfy mediation null criterion under the FFRCISTG model in Figure 1(c). However, the mediation null criterion can be satisfied under either of the following conditions: (i) by incorporating an additional assumption (A5) into the FFRCISTG model in Figure 1(c), or (ii) under the NPSEM-IE model in Figure 1(b), where the IIE and NIE become identifiable as the same quantity, thereby fulfilling the mediation null criterion. Our next task is to determine how to make a choice among these measures based on the strength of the causal models. Figure 3 summarizes the relationships between the five causal models required for identifying these three effects mentioned above: (a) the FFRCISTG in Figure 2(c), (b) NPSEM-IE in Figure 1(b), (c) population FFRCISTG in Figure 2(b), (d) the FFRCISTG in Figure 1(c) with additional Assumption 5, and (e) the FFRCISTG in Figure 1(c) without Assumption 5. The NIE, SIE, and IIE presented in the boxes in Figure 3 are all identified as Formula 1. Therefore, in practical analysis, we focus on identifying the assumptions that allow Formula 1 to correspond to different indirect effect measures. Among these, the IIE is the measure that can be nonparametrically identified as Formula 1 under the weakest assumptions. However, for the IIE to satisfy the mediation null criterion, at least Assumption 4 or Assumption 5 must hold. The assumptions required to identify the SIE are stronger than those for the IIE, as they rely on dismissible component assumptions. However, as shown in Theorem 2, satisfying the mediation null criterion requires at least a cross-world assumption, which necessitates strengthening the assumptions to the FFRCISTG model in Figure 2(c). Under this model, however, the SIE loses its empirical verifiability and requires even stronger assumptions than the NPSEM-IE model in Figure 1(b), which is used to identify the NIE. Thus, while dismissible component assumptions may, to some extent, limit the occurrence of scenarios in which the SIE lacks a mediational interpretation, they do not fully ensure that it satisfies the mediation null criterion. In conclusion, regardless of identifiability, the NIE satisfies the mediation null criterion without requiring any additional assumptions. Moreover, the SIE and IIE can only satisfy the mediation null criterion when they are identified as equal to the NIE. This makes the NIE the most standard measure of indirect effect. On the other hand, if intermediate confounders exist, all models in Figure 3 fail; this case will be addressed in the next section.

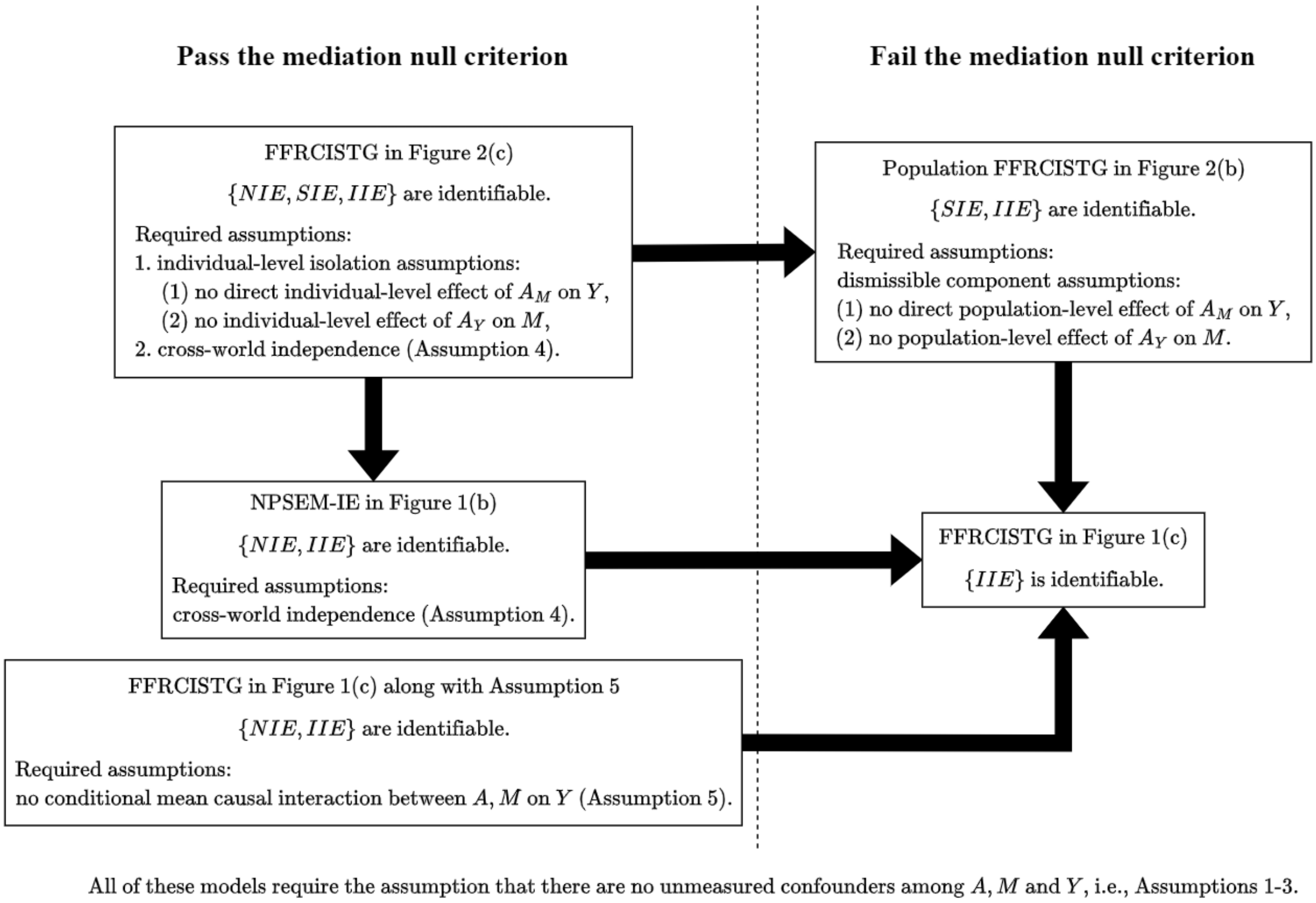

**Figure 3.** The relationship between causal models that used to identify the NIE, IIE, and SIE in the absence of intermediate confounders. In this figure, if there are edges pointing from one model to another, it means the former model implying the latter.

# 3. CAUSAL MEDIATION ANALYSIS WITH INTERMEDIATE CONFOUNDERS

## 3.1 Mediational interpretation for NIE and IIE in the presence of intermediate confounders

In the presence of intermediate confounders, all causal models mentioned in the previous section are no longer applicable. In this section, we aim to clarify that in this situation, (a) NIE is the only measure among the three that both satisfies the mediation null criterion and can be identified, (b) IIE does not meet the mediation null criterion, which corrects a theorem proposed by Miles [26], and (c) the separable effect method cannot even target the indirect effect. VanderWeele, Vansteelandt [4] have proposed that the IIE can still be identified under Figure 4(b), i.e., the FFRCISTG associated with Figure 4(a), even in the presence of unmeasured $L$-$M$ or $L$-$Y$ confounders, but not when both are present. The model implies Assumptions 1, 2, and

**Assumption 10.**

$$Y(a', l', m') \perp\!\!\!\perp M \,\big|\, A = a', L = l', C, \qquad \forall a' \in \Omega_A, l' \in \Omega_L, m' \in \Omega_M,$$

then IIE and IDE can be identified as

**Formula 3.** $IIE \overset{id.}{=} Q^L(1,1) - Q^L(1,0),$

**Formula 4.** $IDE \overset{id.}{=} Q^L(1,0) - Q^L(0,0),$

respectively, where $Q^L(a', a'') = E\left[E\left\{\int_l E(Y|A = a', L = l, M, C)\mathrm{d}F_{L|A,C}(l|a', C) \,\middle|A = a'', C\right\}\right]$. Miles [26] pointed out that IIE does not have a mediational interpretation under the FFRCISTG assumptions in Figure 4(b); but IIE could meet the mediation null criterion by adding Assumption 5, as the identification results would align with NIE. However, we showed that there are flaws in Miles's proof in our Supplementary Material B, where a counterexample reveals that:

**Theorem 3.** *The IIE,* $E\{Y(1, G_C(1)) - Y(1, G_C(0))\}$, *does not have a mediation null criterion under the FFRCISTG in Figure 4(b) along with Assumption 5.*

Since Miles's identification result for $NIE$ is also incorrect, we are going to present another way of identifying NIE. Under Assumption 5, as Miles mentioned, which implies the reference interaction [29] $E\left[\sum_{m' \in \Omega_M}\{Y(1, m') - Y(0, m') - Y(1, m'') + Y(0, m'')\}I(M(0) = m')\right]$ would be zero for all $m'' \in \Omega_M$, such that $NDE = CDE(m')$ and $NIE = PE(m') = TE - CDE(m')$ for all $m' \in \Omega_M$. Thus, under Figure 4(b) and Assumption 5, $TE$ can be identified as $Q^{TE}(1) - Q^{TE}(0)$, where

$$Q^{TE}(a') \equiv E\left[\int_{l'}\int_{m'} E\{Y|M = m', L = l', A = a', C\}dF_{M|L,A,C}(m'|l', a', C)dF_{L|A,C}(l'|a', C)\right],$$

and $NDE = CDE(m')$ can be identified as $Q^{CDE}(1, m') - Q^{CDE}(0, m')$, where

$$Q^{CDE}(a', m') \equiv E\left[\int_{l'} E[Y|M = m', L = l', A = a', C] dF_{L|A,C}(l'|a', C)\right].$$

Then $NIE$ is identified as

**Formula 5.**

$$Q^{TE}(1) - Q^{TE}(0) - Q^{CDE}(1, m') + Q^{CDE}(0, m'), \qquad \text{for some } m' \in \Omega_M.$$

It should be emphasized here that the choice of $m'$ should satisfy the positivity assumption that $P(M = m'|L = l', A = a', C = c') > 0$ whenever $P(L = l', A = a', C = c') > 0$. Next, we will explore the methods developed for the separable effect in the presence of intermediate confounders.

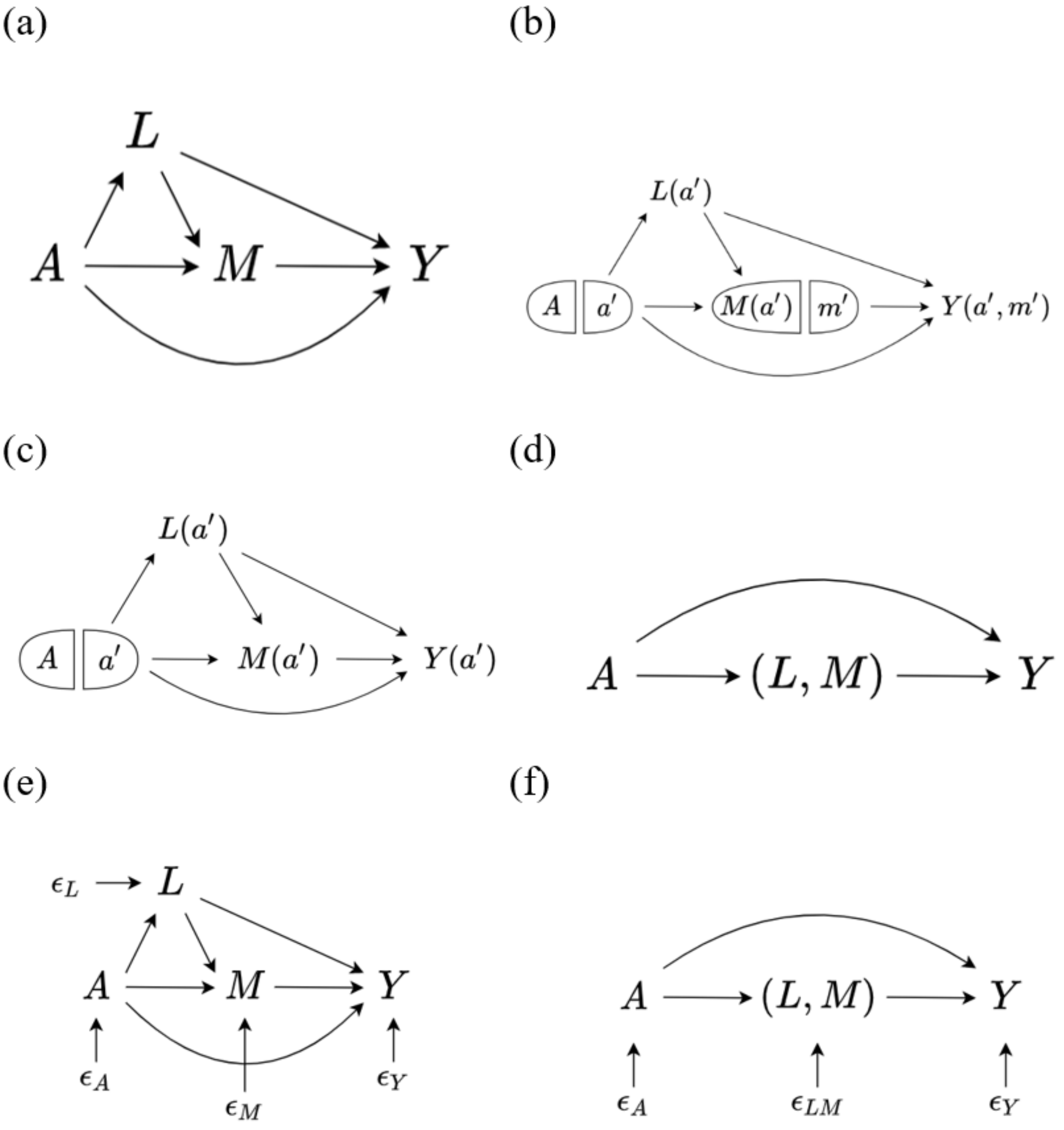

**Figure 4.** (a) A causal DAG for mediation analysis in the presence of $M$-$Y$ confounders $L$ affected by $A$; (b) a template of SWIG representing the FFRCISTG associated with Figure 4(a); (c) another template of SWIG, representing the same model as in Figure 4(b), but with another intervention regime; (d) a causal DAG for mediation analysis considering both $M$ and $Y$ as mediators; (e) an NPSEM-IE associated with Figure 4(a) where $\epsilon_A \perp\!\!\!\perp \epsilon_L \perp\!\!\!\perp \epsilon_M \perp\!\!\!\perp \epsilon_Y$; (f) an NPSEM-IE associated with Figure 4(d) where $\epsilon_A \perp\!\!\!\perp \epsilon_{LM} \perp\!\!\!\perp \epsilon_Y$. For sake of simplicity, the measurable baseline confounders $C$ among $A$, $M$ and $Y$ are not shown in the figures.

### 3.2 No mediational interpretation of separable effects in the presence of intermediate confounders

In the presence of intermediate confounders, the total effect can be decomposed into four pathways as shown in Figure 5: (a) the effect of exposure on the outcome without going through intermediate confounders and the mediator ($A \to Y$), (b) the effect of exposure on the outcome through intermediate confounders but not through the mediator ($A \to L \to Y$), (c) the effect of exposure on the outcome sequentially through intermediate confounders and then the mediator ($A \to L \to M \to Y$), and (d) the effect of exposure on the outcome solely through the mediator ($A \to M \to Y$). According to the definition of mediation, a valid indirect effect should capture both pathway (c) and (d). Figure 5 illustrates the effect pathways captured by $NIE$, $IIE$, and the separable effect method which will be introduced in this subsection. While both $NIE$ and $IIE$ still target the indirect effect, however, the presence of intermediate confounders evidently violates both the individual-level isolation assumptions and dismissible component assumptions, resulting in the separable effect method being unable to establish a causal estimand for a valid indirect effect. Robins, Richardson [8] proposed three strategies by making a strong assumption in Figure 4(a), based on the causal structures shown in Figure 6(a)-(c), to partially address the problem of effect decomposition, but this changed the original causal question. Figures 6(a)-(c) and the extension on population FFRCISTG as shown in Figure 6(d)-(f) further assumes the effect of $A$ on $L$ is completely mediated by $A_M$, $A_Y$ and $A_L$, respectively. Since Figure 6(a)-(c) are submodels of Figure 4(a), all the population FFRCISTGs associated with Figure 6(a)-(c) imply the population FFRCISTG associated with Figure 4(a). As shown in Figure 4(b)-(c), the population FFRCISTG associated with Figure 4(a) implies the Supplementary Assumptions 1-3 in Supplementary Material C.1, which indicate that there are no unmeasured confounders simultaneously affect exposure and either the intermediate confounders, mediator, or outcome. Then the same holds for the population FFRCISTG in Figures 6(d)-(f). Under the model shown in Figure 6(d), separable direct and indirect effects can be identified as

$$SDE \overset{id.}{=} Q_1(1,0) - Q_1(0,0),$$

$$SIE \overset{id.}{=} Q_1(1,1) - Q_1(1,0),$$

where $Q_1(a', a'') \equiv E[E\{E(Y|A = a', L, M, C)|A = a'', C\}]$. The dismissible component assumptions implied by the models corresponding to each decomposition strategies are listed in Supplementary Material C.1. For another decomposition strategy, under the population FFRCISTG in Figure 6(e), separable effects can be identified as

$$SDE \overset{id.}{=} Q_2(1,0) - Q_2(0,0),$$

$$SIE \overset{id.}{=} Q_2(1,1) - Q_2(1,0),$$

where $Q_2(a', a'') \equiv E[E\{E\{E(Y|A = a', L, M, C)|A = a'', L, C\}|A = a', C\}]$. The definitions of the separable effects for these two decomposition strategies are the same, with the difference being that $A_M$ and $A_Y$ must conform to their respective causal DAGs. In Figure 6(c), Robins, Richardson [8] further assume that $A$ consists of three elements $\{A_L^*, A_M^*, A_Y^*\}$, which is a submodel of Figure 6(a) if $A_M \equiv \{A_L^*, A_M^*\}$ and $A_Y \equiv A_Y^*$, and is a submodel of Figure 6(b) if $A_M \equiv A_M^*$ and $A_Y \equiv \{A_L^*, A_Y^*\}$, while they did not formally present a definition of decomposed effects. Let $V_{SL}(a', a'', a''')$ is the abbreviation of counterfactual $V(a_Y^* = a', a_L^* = a'', a_M^* = a''')$ for $V \in \{L, M, Y\}$. Under the population FFRCISTG associated with Figure 6(c), which depicted in Figure 6(f), we define and identify the separable effects as

$$SE_{\text{direct}} \equiv E\{Y_{SL}(1,0,0)\} - E\{Y_{SL}(0,0,0)\} \overset{id.}{=} Q_3(1,0,0) - Q_3(0,0,0),$$

$$SE_L \equiv E\{Y_{SL}(1,1,1)\} - E\{Y_{SL}(1,0,1)\} \overset{id.}{=} Q_3(1,1,1) - Q_3(1,0,1),$$

$$SE_M \equiv E\{Y_{SL}(1,0,1)\} - E\{Y_{SL}(1,0,0)\} \overset{id.}{=} Q_3(1,0,1) - Q_3(1,0,0),$$

where $Q_3(a', a'', a''') \equiv E[E\{E\{E(Y|A = a', L, M, C)|A = a''', L, C\}|A = a'', C\}]$. If Figure 6(c) serves as a submodel for Figure 6(a), then the $SIE$ under Figure 6(a) is essentially equal to $SE_L + SE_M$ under Figure 6(c). Alternatively, if Figure 6(c) is a submodel of Figure 6(b), then the $SIE$ under Figure 6(b) is equal to $SE_M$ under Figure 6(c). Figure 5 presents the decomposition pathways available for the separable effects, emphasizing that the separable effect method falls short in capturing the indirect effect. However, these decomposition strategies have already been developed by VanderWeele, Vansteelandt, and Robins.[4] We will discuss the relationship between separable effects and previous methods in the next subsection.

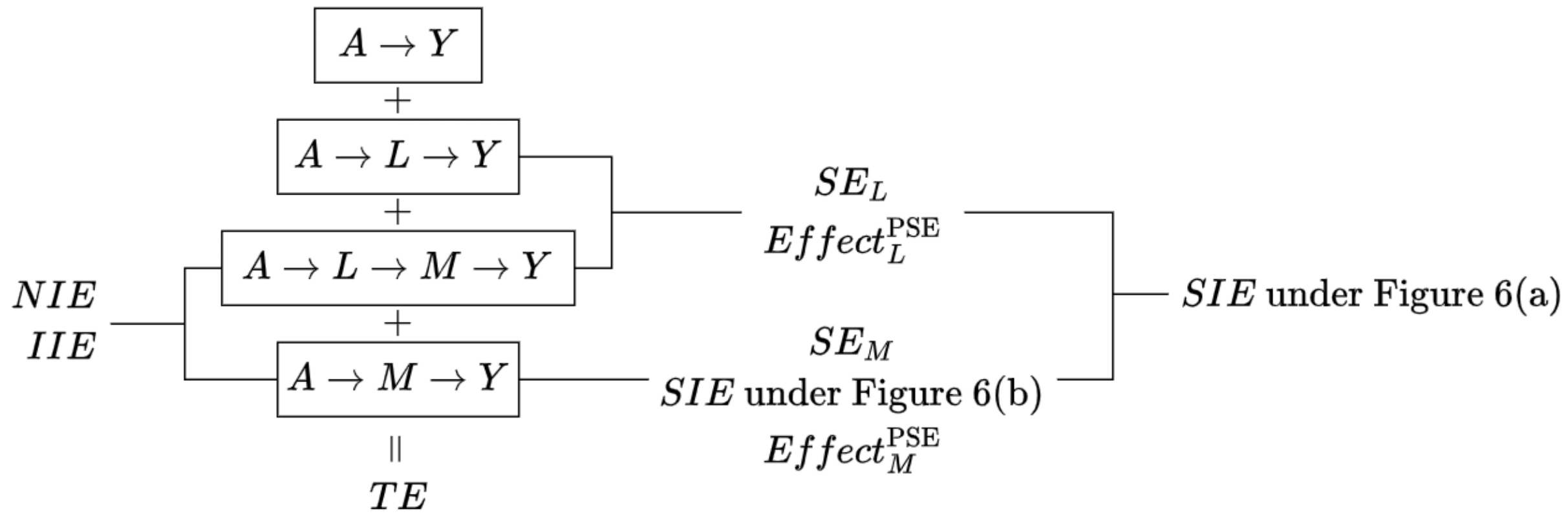

**Figure 5.** The decomposed pathways of the separable effect method and the differences from the NIE and IIE.

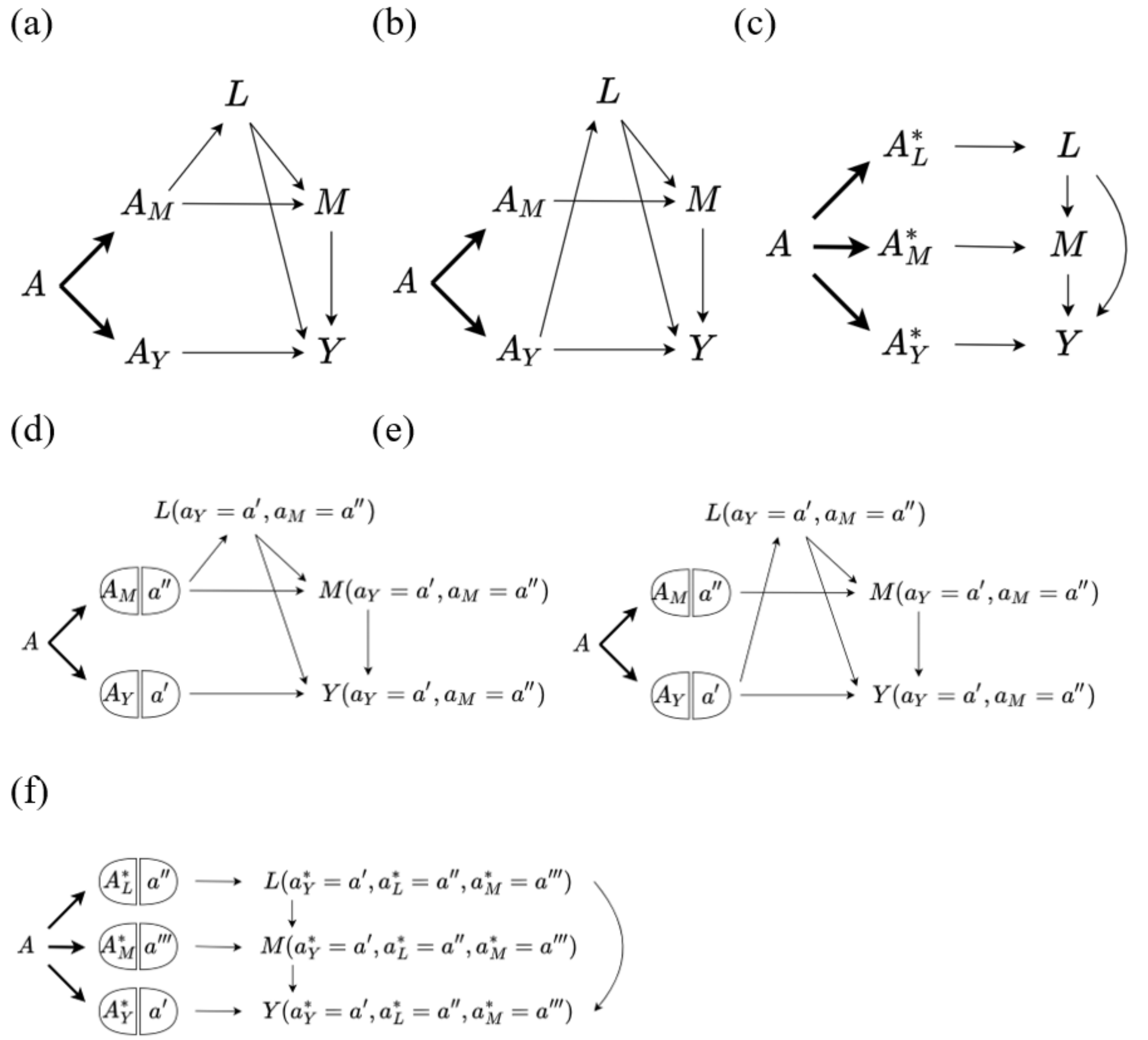

**Figure 6.** (a)-(c) Extended DAGs for mediation analysis in the presence of $M$-$Y$ confounders $L$ affected by $A$, where the thick edges represent that $A_M$, $A_Y$, $A_L^*$, $A_M^*$, and $A_Y^*$ are completely determined by $A$. The figures illustrate three causal structures that the effect of $A$ on $L$ completely mediated by (a) $A_M$, (b) $A_Y$ and (c) $A_L^*$. (d)-(f) The population SWIG representing the population FFRCISTG associated with Figure 6(a)-(c), respectively. For sake of simplicity, the measurable baseline confounders $C$ among $A$, $M$ and $Y$ is omitted in the figures.

### 3.3 The relationship between effect decompositions of separable effects and VanderWeele's methods

If Figure 6(a)-(c) are interpreted at the individual level, that is, they meet the individual-level isolation assumptions in Supplementary Material C.2, then the separable effect decomposition corresponding to Figure 6(a) and Figure 6(c) respectively align with the approaches of "joint mediators" (JM) and "path-specific effects" (PSE) proposed by VanderWeele, Vansteelandt, and Robins [4]. The first approach considers both $L$ and $M$ as mediators (as shown in Figure 4(d)) and decompose the total effect into the effect associated with either $L$ or $M$ ($A \to M \to Y$, $A \to L \to Y$, and $A \to L \to M \to Y$; $Effect_{LM}^{\mathrm{JM}}$) and the remaining direct effect ($A \to Y$; $Effect_{\mathrm{direct}}^{\mathrm{JM}}$). The second approach decompose the total effect into three parts: (a) the effect associated with intermediate confounders ($A \to L \to M \to Y$ and $A \to L \to Y$; $Effect_{L}^{\mathrm{PSE}}$), referred to as the "intermediate-confounder-specific effect", and (b) the remaining indirect effect ($A \to M \to Y$; $Effect_{M}^{\mathrm{PSE}}$) and (c) direct effect ($A \to Y$; $Effect_{\mathrm{direct}}^{\mathrm{PSE}}$). Let $L(a'), M(a', l'), Y(a', l', m')$ are the abbreviations of counterfactuals $L(a = a'), M(a = a', l = l'), Y(a = a', l = l', m = m')$ , respectively. The decomposed effects of JM approach are defined as follows:

$$Effect_{\mathrm{direct}}^{\mathrm{JM}} \equiv \phi^{\mathrm{JM}}(1,0) - \phi^{\mathrm{JM}}(0,0),$$

$$Effect_{LM}^{\mathrm{JM}} \equiv \phi^{\mathrm{JM}}(1,1) - \phi^{\mathrm{JM}}(1,0),$$

where $\phi^{\mathrm{JM}}(a', a'') \equiv E\{Y(a', L(a''), M(a''))\}$. And the decomposed effects of PSE approach are defined as follows:

$$Effect_{\mathrm{direct}}^{\mathrm{PSE}} \equiv \phi^{\mathrm{PSE}}(1,0,0) - \phi^{\mathrm{PSE}}(0,0,0),$$

$$Effect_{L}^{\mathrm{PSE}} \equiv \phi^{\mathrm{PSE}}(1,1,1) - \phi^{\mathrm{PSE}}(1,0,1),$$

$$Effect_{M}^{\mathrm{PSE}} \equiv \phi^{\mathrm{PSE}}(1,0,1) - \phi^{\mathrm{PSE}}(1,0,0),$$

where $\phi^{\mathrm{PSE}}(a', a'', a''') \equiv E\{Y(a', L(a''), M(a''', L(a'')))\}$ . Moreover, if we treat the intermediate-confounder-specific effect as a component of the direct effect, then the decomposition strategy of the $SIE$ and $SDE$ under Figure 6(b) corresponds to that of $Effect_{M}^{\mathrm{PSE}}$ and $Effect_{\mathrm{direct}}^{\mathrm{PSE}} + Effect_{L}^{\mathrm{PSE}}$; again, they are equivalent if the individual-level isolation assumptions in Supplementary Material C.2 are met. Furthermore, all the identification results for these three decomposition strategies are identical to the corresponding results of the separable effect method, despite requiring different causal models.

Figure 7 organizes the relationships between the causal models that can identify the causal effects for different decomposition strategies. The assumptions required for identifying separable effects are different from those for the mediation effects defined through nested counterfactuals. The former can be identified under the population FFRCISTG associated with Figures 6(d)-(f), while the latter is under the NPSEM-IE in Figure 4(e)-(f). The independence assumptions implied by these models are listed in Supplementary Material C.1. In fact, a decomposition strategy of separable effect method always has a corresponding approach defined by nested counterfactuals. However, these three decomposition methods of separable effects cannot aim to identify indirect effect. Therefore, under the causal structures of Figures 4(a), the NIE is still the best approach if the assumptions of FFRCISTG in Figure 4(b) and Assumption 5 hold.

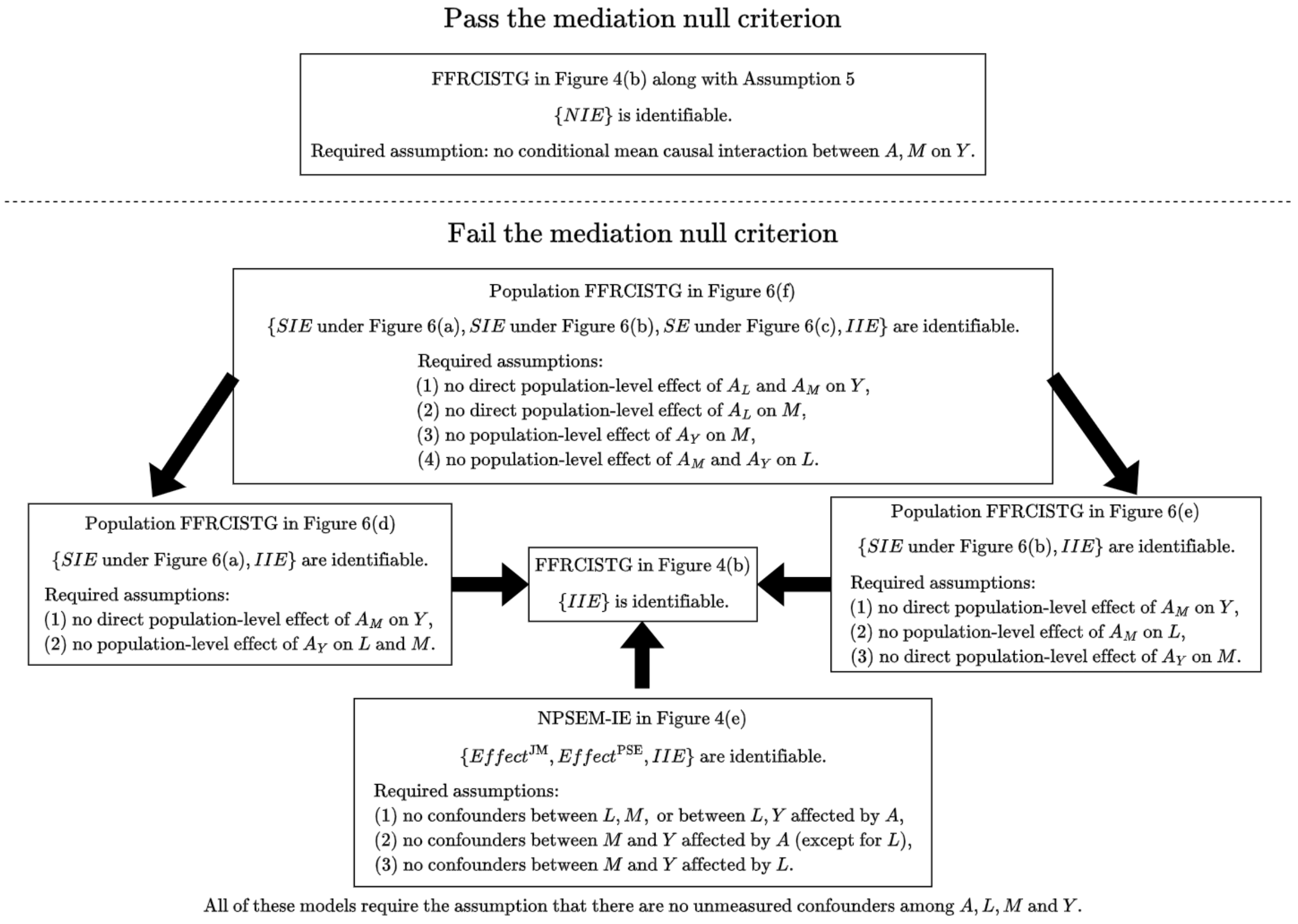

**Figure 7.** The relationship between causal models that used to identify the effects that use different decomposition strategies. In this figure, if there are edges pointing from one model to another, it means the former model implying the latter.

## 4. INTEGRATED FRAMEWORK FOR PRACTICAL ANALYSIS

We further propose a working flowchart for practical analysis as shown in Figure 8, which summarizes all the causal models introduced in this article. We now use the data analysis from VanderWeele, Vansteelandt [4] to illustrate how to use and interpret the flowchart in Figure 8 in practical analysis. In this example, our goal is to understand the indirect effect of prenatal care ($A$: 1 = adequate, 0 = inadequate) through the occurrence of preeclampsia ($M$: 1 = preeclampsia, 0 = no preeclampsia) on preterm birth ($Y$: 1 = preterm, 0 = not preterm), as well as the direct effect of prenatal care on preterm birth. Evidently, maternal smoking behavior ($L$: 1 = smoker, 0 = nonsmoker) is an intermediate confounder (decision #1) because adequate prenatal care should reduce the frequency of smoking during pregnancy. In addition, many studies believe that smoking has an inhibitory effect on preeclampsia [30] and simultaneously increases the risk of preterm birth [31] directly. According to the flowchart in Figure 8, when focusing on the indirect effect (decision #7), according to decision #8, we need to ask whether there are unmeasured exposure-mediator (Assumption 1), exposure-outcome (Assumption 2), mediator-outcome confounders (Assumption 10)? If these conditions are met, then we proceed to decision #9: is there a mean causal interaction between $A$, $M$ on $Y$ conditional on $M(0)$ and $C$ (Assumption 5)? If the researcher believes there is no such interaction, then NIE can be identified as Formula 5. If there is an interaction, then only the interpretation of IIE can be used which is identified as Formula 3. (Assumption 10) may also be violated due to the simultaneous presence of unmeasured $L$-$M$ confounders and unmeasured $L$-$Y$ confounders, then at decision #8, the only choice is to employ the decomposition strategies associated with $Effect^{\mathrm{JM}}$ and $SIE$ under Figure 6(a), which cannot identify the indirect effect.

On the other hand, if we are also interested in the effects related to smoking, and do not request for identifying the indirect effect, then we can choose the decomposition strategies associated with $Effect^{\mathrm{PSE}}$ and $SE$ under Figure 6(c) at decision #10, and they postulate different causal models. In this example, we can reasonably assume that prenatal care ($A$) can be divided into “smoking behavior management and health education ($A_L$)”, “early pre-eclampsia risk assessment and prevention ($A_M$)” and “other prenatal care measures ($A_Y$)” and that these satisfy the FFRCISTG in Figure 6(f). VanderWeele, Vansteelandt [4] have already identified the path-specific effects under the assumptions of NPSEM-IE in Figure 4(e), and we therefore use their results to interpret separable effects under Figure 6(c), because the identifying formulas are identical. The results show that appropriate smoking behavior management and health education ($A_L$) can reduce the odds of preterm birth by 0.7%. Meanwhile, appropriate early pre-eclampsia risk assessment and prevention ($A_M$) can increase the odds of preterm birth by 0.06%, although this result is not significant. One literature[32] has shown that low-dose aspirin reduces the incidence of preeclampsia in high-risk women, but prenatal care in the collected dataset may not have included this measure to prevent preeclampsia. Therefore, the effect of prenatal care in reducing preterm birth through pre-eclampsia but not through smoking is not significant. The remaining direct effect can be interpreted as the impact of other prenatal care measures ($A_Y$) on preterm birth, which reduced the odds of preterm birth by 54.3%.

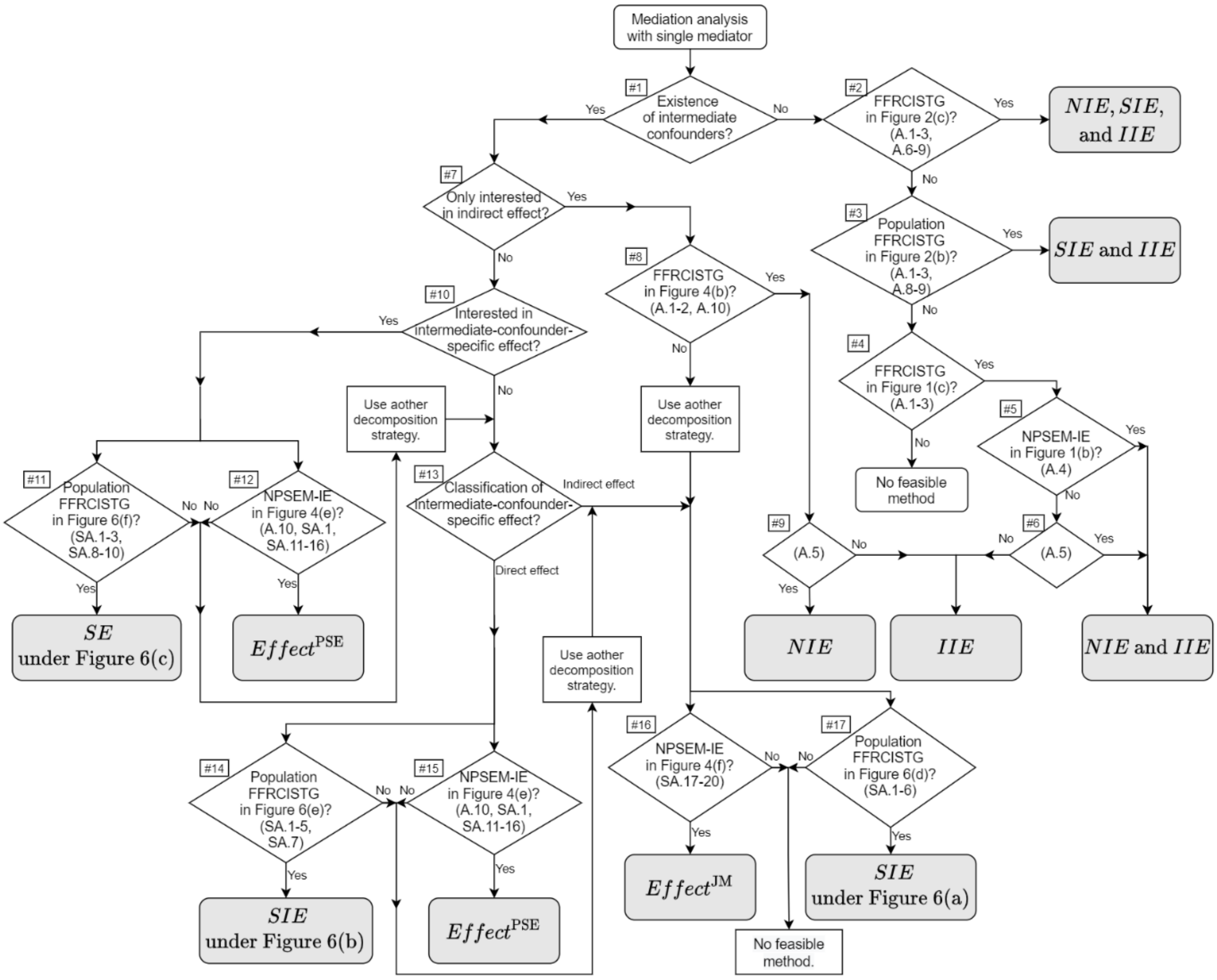

**Figure 8.** The flowchart for practical analysis. This allows for the selection of appropriate mediation effect measures based on the purpose of the analysis and the assumptions made. In this figure, Assumption $i$ and Supplementary Assumption $i$ are abbreviated as A.$i$ and SA.$i$, respectively, for $i = 1,2,3$ ....

# 5. DISCUSSION

In this article, we discuss several causal models used to identify the NIE, IIE, and SIE. We also evaluate these three indirect effect measures using the mediation null criterion proposed by Miles [26]. In particular, we prove that the SIE lacks a mediational interpretation in the absence of intermediate confounders unless it relies on assumptions stronger than those required for NPSEM-IE. Since the identification results for the NIE, IIE, and SIE are equivalent in the absence of intermediate confounders, according to Figure 3, one can always adopt conservative assumptions and interpret the results as IIE. If stronger assumptions are believed to hold based on background knowledge, one may further adopt the interpretation of NIE or SIE. However, when the SIE and IIE are not identifiable as the NIE, neither can satisfy the mediation null criterion under the existing models.

When intermediate confounders are present, we demonstrate that the IIE lacks a mediational interpretation under the FFRCISTG model in Figure 4(b) along with Assumption 5, a conclusion that differs from the claim made by Miles [26]. We also show that the separable effect method fails to capture the indirect effect. Although the alternative path decomposition strategies proposed by Robins, Richardson [8] can be applied, the resulting effect of interest no longer has a mediational interpretation. This study further establishes that all path decomposition strategies for the separable effect method correspond to causal effects defined by nested counterfactuals. The key distinction is that the latter approach requires cross-world assumptions, whereas the former does not, instead relying on specific causal structures. Nonetheless, in this case, the NIE remains identifiable under a no-interaction assumption, and we present a new identification result.

In summary, this article highlights the challenges associated with the SIE in causal mediation analysis, clarifies its relationship with the NIE and IIE, and provides an integrated framework for practical analysis. Since Robins and Richardson [5] introduced the separable effect method, researchers conducting causal mediation analysis have gained additional options for defining and interpreting a mediation parameter of interest. This study demonstrates that the SIE does not satisfy the mediation null criterion and, consequently, also fails to meet the two stricter criteria proposed by Miles [26], which are beyond the scope of this article. Nevertheless, due to its empirical verifiability and intuitive interpretation, the SIE remains valuable in practice. Future research could develop more flexible criteria to justify and evaluate its mediational interpretation. Furthermore, we establish that the NIE is the only indirect effect measure among the three that satisfies the mediation null criterion. However, its identification assumptions are inherently untestable, underscoring the fundamental trade-off between mediational interpretability and the falsifiability of identification assumptions. A key direction for future research in causal mediation analysis is to explore less restrictive assumptions that strike a balance between these two critical aspects.

## FUNDING

This work was supported by the National Science and Technology Council of Taiwan [Grant No. MOST 111-2628-M-A49-004-MY3].

## COMPETING INTERESTS

The authors have no relevant financial or non-financial interests to disclose.

## AUTHOR CONTRIBUTIONS

Yan-Lin Chen drafted the manuscript. Sheng-Hsuan Lin supervised the research. All authors read and approved the final manuscript.

## DATA AVAILABILITY STATEMENT

Data sharing is not applicable to this article as no new data were created or analyzed in this study.

## REFERENCES

1. VanderWeele TJ. Explanation in causal inference: methods for mediation and interaction.: Oxford University Press; 2015.

2. Robins JM, Greenland S. Identifiability and exchangeability for direct and indirect effects. Epidemiology 1992;3(2):143-55. doi:10.1097/00001648-199203000-00013

3. Pearl J, editor. Direct and indirect effects. Proceedings of the Seventeenth conference on Uncertainty in artificial intelligence; 2001; San Francisco, CA, USA: Morgan kaufmann publishers Inc.

4. VanderWeele TJ, Vansteelandt S, Robins JM. Effect decomposition in the presence of an exposure-induced mediator-outcome confounder. Epidemiology. 2014;25(2):300-6. doi:10.1097/ede.0000000000000034

5. Robins JM, Richardson TS. Alternative graphical causal models and the identification of direct effects. In: Shrout PE, Keyes KM, Ornstein K, editors. Causality and psychopathology: finding the determinants of disorders and their cures. 1st ed: Oxford University Press; 2010. p. 103-58.

6. Stensrud MJ, Young JG, Didelez V, Robins JM, Hernán MA. Separable effects for causal inference in the presence of competing events. Journal of the American Statistical Association. 2022;117(537):175-83. doi:10.1080/01621459.2020.1765783

7. Stensrud MJ, Robins JM, Sarvet A, Tchetgen Tchetgen EJ, Young JG. Conditional separable effects. Journal of the American Statistical Association. 2023;118(544):2671-83. doi:10.1080/01621459.2022.2071276

8. Robins JM, Richardson TS, Shpitser I. An interventionist approach to mediation analysis. Probabilistic and Causal Inference: The Works of Judea Pearl: Association for Computing Machinery; 2022. p. 713–64.

9. Didelez V. Defining causal mediation with a longitudinal mediator and a survival outcome. Lifetime Data Anal. 2019;25(4):593-610. doi:10.1007/s10985-018-9449-0

10. Aalen OO, Stensrud MJ, Didelez V, Daniel R, Røysland K, Strohmaier S. Time-dependent mediators in survival analysis: Modeling direct and indirect effects with the additive hazards model. Biometrical Journal. 2020;62(3):532-49. doi:10.1002/bimj.201800263

11. Holland PW. Statistics and causal inference. Journal of the American Statistical Association. 1986;81(396):945-60. doi:10.2307/2289064

12. Pearl J. Causality: Models, Reasoning, and Inference. 2nd ed. New York: Cambridge University Press; 2009.

13. VanderWeele TJ, Robinson WR. Rejoinder: How to Reduce Racial Disparities?: Upon What to Intervene? Epidemiology. 2014;25(4):491-3. doi:10.1097/ede.0000000000000124

14. Geneletti S. Identifying direct and indirect effects in a non-counterfactual framework. Journal of the Royal Statistical Society Series B. 2007;69(2):199-215. doi:10.1111/j.1467-9868.2007.00584.x

15. Didelez V, Dawid AP, Geneletti S, editors. Direct and indirect effects of sequential

treatments. Proceedings of the Twenty-Second Conference on Uncertainty in Artificial Intelligence; 2006.

16. Moreno-Betancur M, Carlin JB. Understanding interventional effects: A more natural approach to mediation analysis? Epidemiology. 2018;29(5):614-7. doi:10.1097/ede.0000000000000866

17. Nguyen TQ, Schmid I, Stuart EA. Clarifying causal mediation analysis for the applied researcher: Defining effects based on what we want to learn. Psychological Methods. 2021;26(2):255.

18. Robins J. A new approach to causal inference in mortality studies with a sustained exposure period—application to control of the healthy worker survivor effect. Mathematical modelling. 1986;7(9-12):1393-512.

19. Greenland S, Pearl J, Robins JM. Causal diagrams for epidemiologic research. Epidemiology. 1999;10(1):37-48.

20. Richardson TS, Robins JM. Single world intervention graphs (SWIGs): A unification of the counterfactual and graphical approaches to causality. Center for the Statistics and the Social Sciences, University of Washington Series. Working Paper. 2013;128(30):2013.

21. Pearl J. Causal diagrams for empirical research. Biometrika. 1995;82(4):669-88. doi:10.2307/2337329

22. VanderWeele TJ, Vansteelandt S. Conceptual issues concerning mediation, interventions and composition. Statistics and Its Interface. 2009;2:457-68.

23. VanderWeele TJ. Concerning the consistency assumption in causal inference. Epidemiology. 2009;20(6):880-3. doi:10.1097/EDE.0b013e3181bd5638

24. Hernán MA, Robins JM. Causal Inference: What If. Boca Raton: Chapman & Hall/CRC; 2020.

25. Andrews RM, Didelez V. Insights into the cross-world independence assumption of causal mediation analysis. Epidemiology. 2021;32(2):209-19. doi:10.1097/ede.0000000000001313

26. Miles CH. On the causal interpretation of randomised interventional indirect effects. Journal of the Royal Statistical Society Series B: Statistical Methodology. 2023;85(4):1154-72. doi:10.1093/jrsssb/qkad066

27. Petersen ML, Sinisi SE, van der Laan MJ. Estimation of direct causal effects.

Epidemiology. 2006;17(3):276-84. doi:10.1097/01.ede.0000208475.99429.2d

28. Stensrud MJ, Hernán MA, Tchetgen Tchetgen EJ, Robins JM, Didelez V, Young JG. A generalized theory of separable effects in competing event settings. Lifetime Data Anal. 2021;27:588–631. doi:10.1007/s10985-021-09530-8

29. VanderWeele TJ. A unification of mediation and interaction: A 4-way decomposition. Epidemiology. 2014;25(5):749-61. doi:10.1097/ede.0000000000000121

30. Karumanchi SA, Levine RJ. How does smoking reduce the risk of preeclampsia? Hypertension. 2010;55(5):1100-1. doi:doi:10.1161/HYPERTENSIONAHA.109.148973

31. Wisborg K, Henriksen TB, Hedegaard M, Jergen N. Smoking during pregnancy and preterm birth. BJOG: An International Journal of Obstetrics & Gynaecology. 1996;103(8):800-5. doi:10.1111/j.1471-0528.1996.tb09877.x

32. Rolnik DL, Wright D, Poon LC, et al. Aspirin versus Placebo in Pregnancies at High Risk for Preterm Preeclampsia. New England Journal of Medicine. 2017;377(7):613-22. doi:10.1056/NEJMoa1704559